# Decentralized Identity in Practice: Benchmarking Latency, Cost, and Privacy

Abylay Satybaldy[a,*], Kamil Tylinski[a,b] and Jiahua Xu[a,b]

[a]*Exponential Science, George Town, Cayman Islands*
[b]*University College London, London, United Kingdom*

## ABSTRACT

Decentralized Identifiers (DIDs) are increasingly deployed on distributed ledgers, yet systematic cross-platform evidence on their operational behavior remains limited. We present an empirical benchmarking study of three prominent ledger-based DID methods—Ethereum (did:ethr), Hedera (did:hedera), and XRP Ledger (did:xrpl)—using reference Software Development Kits (SDKs) under a unified experimental setup. We measure latency, transaction cost, and on-chain metadata exposure, normalizing latency by each platform's block or consensus interval and cost by its native value transfer fee. Privacy leakage is quantified using a Metadata-Leakage Score (MLS), an entropy-based measure expressed in bits per operation.

Our results reveal distinct architectural trade-offs. Ethereum enables near-instant, off-chain DID creation, but incurs the highest latency and cost for on-chain lifecycle operations. XRPL delivers deterministic and stable latency with fixed, low fees, yet exhibits higher metadata leakage due to more verbose transaction payloads. Hedera achieves the lowest on-chain latency and low fees with minimal metadata leakage, while occasional variance arises from SDK-side processing and confirmation pipelines.

Overall, the findings show that ledger architecture and SDK workflows play a major role in shaping DID latency, cost, and metadata exposure, complementing the effects of the underlying consensus mechanism. These results provide evidence-based insights to support informed selection and configuration of DID systems under performance and privacy constraints.

## 1. Introduction

Decentralized Identity enables user–controlled identifiers and credentials without centralized authorities. Individuals can own, manage, and selectively disclose credentials [1]. In contrast, conventional identity (government, corporate SSO, OpenID Connect) relies on intermediary databases that create single points of failure and cross–service tracking risks [2, 3].

Crypto wallets partially address this by providing self–generated key pairs, but they lack a standard way to express verifiable claims or bind identifiers to rich attributes [4]. DID frameworks fill this gap via W3C Decentralized Identifiers (DIDs) [5] and Verifiable Credentials (VCs) [6]: a DID resolves to a DID Document (keys, service endpoints, rotation metadata) hosted on decentralized infrastructure (e.g., blockchains or InterPlanetary File System (IPFS)), while credentials are proven with JSON Web Tokens (JWTs) or Linked Data Proofs [7, 8].

As adoption grows, multiple distributed ledger (DLT) platforms have introduced their own DID methods, each with distinct architectures. Hereafter, 'blockchain' and DLT are used interchangeably. For real-world viability, decentralized identity systems must balance responsiveness, cost, and privacy—yet systematic, empirical comparisons of these trade-offs remain scarce.

a.satybaldy@exp.science (A. Satybaldy);
kamil.tylinski.16@ucl.ac.uk (K. Tylinski); jiahua.xu@ucl.ac.uk (J. Xu)
ORCID(s): 0000-0002-7735-4902 (A. Satybaldy); 0009-0005-4611-2336
(K. Tylinski); 0000-0002-3993-5263 (J. Xu)

To address this gap, we benchmark three widely used DID methods—Ethereum (did:ethr), Hedera (did:hedera), and XRP Ledger (did:xrpl)—using reference SDKs and a uniform benchmarking setup. We evaluate (i) latency, (ii) transaction cost, and (iii) on–chain metadata leakage.

The study is guided by three research questions:

- **RQ1 (Performance):** What are the absolute and relative latency, cost, and metadata–leakage characteristics of core DID operations under standardized experimental conditions?

- **RQ2 (Attribution):** Which architectural design choices and SDK workflows account for the performance and privacy outcomes observed in RQ1?

- **RQ3 (Trade–offs):** How do the evaluated platforms compare in balancing responsiveness, cost efficiency, and privacy for practical DID deployments?

Our goal is to move beyond qualitative comparisons toward a structured, data–driven evaluation that supports evidence-based system selection and design decisions. This work provides an empirical, system-level analysis of how existing DID design choices translate into concrete latency, cost, and privacy properties. Our contribution is threefold: (i) a unified and reproducible benchmarking framework that isolates DID-specific overhead across heterogeneous ledger architectures; (ii) empirical identification of non-obvious drivers of performance and cost; and (iii) a quantitative analysis of on-chain metadata exposure that reveals key privacy leakage drivers. Together, these findings translate raw measurements into evidence-based design insights for DID system selection and configuration.

The remainder is organized as follows: Section 2 surveys prior work; Section 3 outlines DID architectures and reference implementations; Section 4 details metrics and experimental setup; Section 5 reports results and analysis; Section 7 concludes.

## 2. Related Work

Several surveys have mapped the evolution of decentralized and self–sovereign identity (SSI) systems. Dunphy and Petitcolas [9] provided an early analysis of blockchain–based identity management, while Mühle [10] outlined the core SSI architecture. More recently, Krul et al. [11] systematized knowledge across SSI implementations by comparing trust assumptions and architectural requirements. Other comparative works, such as Satybaldy et al. [12] and Dib et al. [13], proposed evaluation frameworks and mapped design components of SSI platforms like uPort and Hyperledger Indy, but relied mainly on documentation rather than empirical testing.

These studies clarify conceptual models yet remain qualitative. To date, no quantitative, cross–platform benchmarking of DID systems exists that measures latency, cost, or metadata leakage of core operations under controlled conditions. This absence of empirical evidence limits their value for system designers assessing trade–offs across ledgers.

In parallel, performance analyses of distributed ledgers have been conducted independently of identity use cases— e.g., Amherd et al. [14] on Hedera Hashgraph, and others on Ethereum and XRPL [15, 16, 17]. These focus on generic throughput or consensus efficiency rather than DID functionality.

Our work addresses this gap by providing the first cross–platform, data–driven comparison of DID reference implementations on Ethereum, Hedera, and XRPL. By benchmarking key DID operations and linking performance to architectural design, we complement prior conceptual surveys with reproducible empirical evidence to guide decentralized identity deployment.

## 3. Background

### 3.1. Decentralized Identifiers (DIDs)

DIDs are W3C–standardized identifiers that enable verifiable, self–managed digital identity without centralized registries [5]. Each DID follows a uniform resource identifier (URI) syntax:

```
did:<method>:<method-specific-identifier>
```

Here, `did` is the URI scheme, `method` specifies the implementation (e.g., `did:ion`, `did:web`, `did:ethr`), and the `identifier` is a unique string, often derived from a public key or hash of cryptographic material. DIDs are globally unique, cryptographically verifiable, and resolvable to structured documents that describe associated keys and endpoints.

### 3.2. DID Document

A DID resolves to a *DID Document*, a JSON–based structure containing the information required to authenticate the DID subject and interact with associated services [18]. Key components include verification methods, service endpoints, and controller metadata. DID Documents can be serialized in JSON or JSON–LD for semantic interoperability. Storage models vary by method:

- On–chain: Entire DID Document stored directly on a DLT.

- Off–chain: Only references or hashes stored on–chain; the document itself resides in IPFS or another distributed storage layer.

The example below presents a minimal W3C DID v1.0– compliant DID Document. The `@context` defines the JSON–LD vocabulary and `id` denotes the DID. A single `verificationMethod` specifies a public key, which is referenced by `authentication` and `assertionMethod` to indicate its permitted uses. The `service` entry exposes a Verifiable Credential endpoint. In practice, DID Documents may include additional relationships and support key rotation through updates to verification methods.

```
{
  "@context": "https://www.w3.org/ns/did/v1",
  "id": "did:example:123456789abcdefghi",
  "verificationMethod": [
    {
      "id": "did:example:123456789abcdefghi#key-1",
      "type": "Ed25519VerificationKey2018",
      "controller": "did:example:123456789abcdefghi",
      "publicKeyBase58": "GfH345jHK...asd67X"
    }
  ],
  "authentication": [
    "did:example:123456789abcdefghi#key-1"
  ],
  "assertionMethod": [
    "did:example:123456789abcdefghi#key-1"
  ],
  "service": [
    {
      "id": "did:example:123456789abcdefghi#vcs",
      "type": "VerifiableCredentialService",
      "serviceEndpoint": "https://example.com/vcs"
    }
  ]
}
```

### 3.3. DID Operations

Core DID operations define how identifiers and their documents are created, updated, and managed [5], as summarized in Table 1. These operations form the lifecycle of a DID and are implemented differently across methods (e.g., smart contract calls, ledger transactions, or consensus messages).

### 3.4. DID Method

A DID Method defines how DIDs are *created*, *resolved*, *updated*, and *deactivated* on a given network. Each method specifies (i) where DID Documents live (on-chain vs. off-chain), (ii) how state changes are recorded (smart contract, ledger-native object, or event stream), and (iii) how resolution reconstructs the current DID state. DID resolution is performed by method-specific *DID Resolvers* that interpret the method and fetch or reconstruct the DID Document from the declared storage backend.

**Table 1**
DID Operations

| Operation | Description |
|---|---|
| **Create** | Registers a new DID and publishes its initial DID Document. |
| **Resolve** | Retrieves the current DID Document from the corresponding storage backend (on–chain or off–chain). |
| **Update** | Modifies the DID Document to reflect new keys, authentication methods, or service end-points. |
| **Revoke** | Removes or deactivates specific keys or at-tributes from the DID Document to prevent further use. |
| **Deactivate** | Permanently decommissions a DID, making it non–resolvable and invalid for future opera-tions. |

**Table 2**
DID SDKs across DLTs

| DLT | DID Method | SDK | Vers. | Language | Chain |
|---|---|---|---|---|---|
| Ethereum | did:ethr [19] | Ethr-DID SDK [20] | 2.1.2 | JavaScript | Sepolia |
| Hedera | did:hedera [21, 22] | DID SDK [23] | 0.1.1 | JavaScript | Hedera Testnet |
| XRPL | did:xrpl [24] | XRPL SDK [25] | 2.4.1 | JavaScript | XRPL Testnet |

## 4. Methodology

We develop a systematic benchmarking framework to evaluate DID reference implementations across multiple distributed ledger platforms. The framework measures both core DID operations and complete identity workflows, covering the essential lifecycle events that define decentralized identity management.

### 4.1. Selection Rationale

We selected Ethereum, Hedera, and XRPL because they are production-grade DID methods that represent three distinct architectural approaches to on-chain identity: contract-based (Ethereum), event-stream–based (Hedera), and ledger-native (XRPL). This provides a meaningful comparative analysis across the dominant design patterns used in current DID systems. Each provides a mature, standards-aligned SDK suitable for reproducible benchmarking. Other DID methods were excluded due to reliance on off-chain storage (e.g., did:web, did:key), lack of stable SDKs, or mis-alignment with our focus on evaluating ledger-level latency, cost, and metadata exposure.

### 4.2. Reference Implementations

We relied on widely used reference implementations available at the time of the study. Hedera does not mandate a single canonical DID method, and multiple community-driven schemes have been proposed and may continue to evolve. For this study, we used a community-proposed DID SDK based on HIP-29 (`did:hedera` [26]) and hosted under the official Hedera GitHub organization. This implementation was selected because, at the time of experimentation, it was the most mature and actively maintained Hedera-compatible DID solution, offering the most comprehensive support for DID operations.

For XRPL, we used the reference DID implementation for the `did:xrpl` method specified in XLS-0040, a community-developed contribution that is officially maintained and documented by the XRP Ledger Foundation. This implementation was selected because, at the time of the study, it was the only publicly specified and supported DID method for XRPL.

For Ethereum, we employed the `did:ethr` method maintained by the Decentralized Identity Foundation (DIF), which is widely adopted within the Ethereum ecosystem.

All three DID methods conform to the W3C DID specification and implement equivalent operations for creation, resolution, update, and deletion. To maintain consistency, all benchmarks were executed with JavaScript-based SDKs, ensuring a uniform environment and tooling across platforms. Table 2 summarizes the DID methods and corresponding SDKs evaluated.

### 4.3. Evaluation Metrics

The benchmarking framework evaluates DID implementations across three key dimensions: performance, cost efficiency, and privacy. Table 3 summarizes the core metrics used in the analysis.

#### 4.3.1. Full Cycle Operation

To capture the end-to-end performance of a realistic DID lifecycle, we introduce a composite metric termed `Full Cycle Operation`. This metric aggregates the measured latency and cost of the core lifecycle actions—Create, Resolve, Update, Revoke, and Delete. The Full Cycle model thus provides a holistic view of cumulative performance and cost, complementing the insights from isolated benchmarks.

#### 4.3.2. Relative latency and cost metrics

To facilitate fair comparison across DLTs with heterogeneous network properties, we normalize latency and cost using platform-specific baselines. For latency, the `relative latency` $L_{\text{rel}}$ is defined as:

$$L_{\text{rel}} = \left( \frac{\bar{L}_{\text{op}}}{\bar{T}_{\text{block}}} \right) \times 100\%$$

**Table 3**
Key Metrics for DID SDK Benchmarking

| Category | Metric | Description |
|---|---|---|
| Performance | Latency | Measures the time elapsed from initiation to completion of each DID operation. |
| Cost Efficiency | On-chain operation cost | Quantifies transaction fees in both native tokens and USD to evaluate economic efficiency. |
| Privacy | On-chain metadata leakage | Quantifies the amount of identifying or fingerprintable information disclosed via on-chain metadata per operation. |

where $\bar{L}_{op}$ is the mean latency of a DID operation, and $\bar{T}_{block}$ is the average block time (or consensus event interval) of the platform [27]. This metric contextualizes responsiveness relative to native block production rates.

For cost, the `relative cost` $C_{rel}$ compares the mean fee of a DID operation $\bar{C}_{op}$ to the cost of a standard token transfer $\bar{C}_{std}$:

$$C_{rel} = \left( \frac{\bar{C}_{op}}{\bar{C}_{std}} \right) \times 100\%$$

*Motivation and Limitations.* Relative metrics allow us to compare heterogeneous ledgers on a common scale despite large differences in block times, fee markets, and transaction models. Normalizing latency by the platform's native consensus interval highlights how much of an operation's delay stems from the DID method itself rather than from the underlying protocol. Similarly, normalizing fees by a standard value transfer exposes the economic overhead introduced specifically by DID operations.

However, relative metrics also have limitations. They abstract away absolute user-perceived performance, and they can downplay the real impact of high mainnet fees or congestion on platforms such as Ethereum. Relative metrics therefore complement, but do not replace, the absolute measurements reported in Figures 4 and 6. Both views are needed for a complete interpretation of system behavior. Moreover, we note that relative normalization is most informative for on-chain operations, whereas for off-chain operations, being inherently independent of consensus delays, it primarily confirms expected performance differences rather than providing additional insight.

### 4.3.3. Metadata–Leakage Score (MLS)

MLS quantifies how much identifying or fingerprintable information is revealed by the *on–chain* metadata of DID operations. Conceptually, MLS computes the Shannon entropy of observable transaction payloads to assess the diversity and uniqueness of metadata content. Expressed in bits *per operation*, higher MLS indicates greater information leakage (stronger fingerprintability), while lower MLS implies more uniform, harder–to–distinguish payloads.

The analysis considers only attributes visible to any blockchain observer (e.g., indexers or mirror nodes) at the moment a DID operation is performed. These include DID–specific fields (identifiers, DID Document parameters such as service endpoints and keys, and cryptographic signatures) and ledger metadata (sequence numbers, block-/ledger numbers, transaction hashes, consensus timestamps). Off–chain signals (e.g., RPC/mirror logs, client IP/TLS metadata, local timestamps) are explicitly excluded.

We evaluate MLS for the three on–chain DID operations—Update, Revoke, and Delete—on Ethereum, Hedera, and XRPL. For each operation and chain, we sample 100 transactions (the same sample used for latency and cost), decode payloads, and aggregate metadata for MLS computation. An example Hedera Update JSON payload is included in the Appendix. Create is excluded from MLS because did:ethr creation is entirely off-chain and produces no on-chain metadata; since MLS measures only ledger-visible information, the Create operation would trivially yield zero and is therefore omitted.

*Motivation and scope.* MLS is designed to quantify the amount of information that an uninvolved public-ledger observer can extract from the metadata emitted by DID lifecycle operations. Our goal is not to provide a complete account of all privacy threats in decentralized identity, but to measure the *fingerprintability* of the on-chain payload itself. In the DID context, even seemingly random or low-entropy fields can aid linkage when repeated across operations; entropy-based metrics therefore provide a principled way to compare how much raw distinguishing signal each method exposes. We adopt a public-ledger passive-observer adversary because it captures the minimum, unpreventable visibility any validator, indexer, or block-explorer has. Stronger adversaries (e.g., wallet-level tracking, cross-protocol correlation, address deanonymization) require off-chain vantage points and auxiliary data that fall outside our system-level benchmarking focus. MLS should thus be interpreted as a lower-bound, method-agnostic measure of on-chain metadata surface, complementary to, but not replacing, contextual or correlation-based privacy analyses.

*Computation.* We proceed in two stages:

1. *Tokenization and entropy calculation.* Each decoded JSON payload is flattened to key–value pairs; each pair is treated as a discrete token. From the empirical distribution $p(x)$ over tokens $x$, we compute Shannon entropy

$$H_{token} = - \sum_x p(x) \log_2 p(x),$$

and normalize by the total token count to obtain average entropy per token.

2. *Aggregation.* We scale by the average token count per transaction to obtain the per–operation MLS:

$$\text{MLS}_{\text{op}} = H_{\text{token}} \times \mathbb{E}[\text{tokens per transaction}].$$

This scaling accounts for differences in payload structure/length across blockchains and operations, enabling a fair per–operation comparison. An MLS of 0 bits/operation indicates fully uniform payloads (indistinguishable to an observer); larger values reflect enough structural or content variability to support fingerprinting.

Employing Shannon entropy follows established practice in privacy/fingerprinting research (e.g., entropy of browser fingerprints [28] and anonymity analyses[29]). Our MLS adapts this information–theoretic approach to the specific case of on–chain DID metadata.

*Adversary model, attack surface, and user contexts.* We evaluate MLS under a public–ledger passive observer model: an adversary who can read any on–chain record and mirror–node/indexer outputs but cannot alter protocol behavior. Capabilities include enumerating DID operations, parsing payload fields, timestamp/sequence correlation, and basic cross–chain/topic joins. Out–of–scope are off–chain vantage points and active manipulation.

*Attack surface.* Tokens considered by MLS are the on–chain fields visible to this observer: DID strings and fragments, service endpoint URIs or hashes, key/verification material identifiers, operation types, transaction/ledger identifiers (hashes, sequence numbers), and consensus timestamps. These are precisely the units tokenized in MLS.

*User contexts.* (i) Holders/wallets originate Update/Revoke/Delete; recurring patterns (e.g., endpoint formats, key–rotation cadence) can enable linkage. (ii) Issuers may publish service endpoints that reveal organizational domains. (iii) Verifiers primarily perform Resolve (read–only, no fee); resolve queries themselves are off–chain and thus outside MLS.

*Actionability.* Within this model, lower MLS means fewer or more uniform tokens per operation, reducing fingerprintability. Practical mitigations include schema normalization and field minimization, hashing or referencing endpoints instead of inlining full URLs, batching/templatizing updates, and avoiding gratuitous custom fields. These mitigations target the same token set MLS measures and thus directly reduce the observable signal.

*Positioning with prior DID/VC privacy work.* Prior work on DID and VC privacy has largely focused on qualitative method comparisons, architectural guidance, or cryptographic guarantees for credential presentation flows, such as unlinkability and selective disclosure in AnonCreds, Idemix, BBS+, and SD–JWT [30, 31, 32, 33]. Standards including the W3C DID Core and the DID Method Rubric articulate privacy considerations at the design and method level, but do not quantify the metadata exposed through ledger interactions [5, 34]. In contrast, MLS targets the *on–chain DID operation layer* and provides per–operation, empirical measurements of publicly observable metadata during Create, Update, Revoke, and Delete events. This bridges a gap between high-level privacy frameworks and cryptographic proofs by offering a chain-specific, operational metric for comparing real-world metadata exposure across DID methods.

## 4.4. Experimental Design and Tools

Benchmarks reflect baseline performance using default SDK configurations, with no platform-specific optimizations. This design ensures fair and reproducible measurements across platforms.

Latency was measured using Node.js high-resolution performance timers, and transaction costs were obtained via ethers.js and blockchain explorer APIs, then converted to USD using contemporaneous market prices. Experiments were executed on an Amazon EC2 c5.4xlarge instance (16 vCPUs, Intel Xeon Platinum @ 3.0 GHz, 32 GiB RAM) running Amazon Linux 2023, providing a consistent execution environment.

The benchmarking experiment was designed as a single, unified evaluation of DID SDKs across three DLT platforms. Each DID SDK was tested with 100 iterations per operation, capturing latency, transaction cost, and on-chain metadata to derive performance, cost-efficiency, and privacy metrics. This controlled setup ensures comparability and isolates the effects of the underlying technology stacks. All reported results are derived from this experimental design.

## 5. Results

We compare core DID operations on three DLT platforms—Ethereum (did:ethr), XRP Ledger (did:xrpl), and Hedera (did:hedera). Our analysis couples empirical measurements with structural examination of each reference SDK's workflows. This combined view quantifies performance while explaining it through architectural choices, providing a comprehensive account of how each platform realizes decentralized identity functionality.

## 5.1. Latency and Design Analysis

We first examine the underlying DID operation mechanisms to contextualize the observed performance differences. Figures 1, 2, and 3 illustrate the sequence diagrams for Create, Resolve, and Update/Revoke/Delete operations, highlighting platform-specific design choices relevant to RQ2. We then present the empirical latency results: Figures 4 and 5 report absolute and relative latencies for the five core DID operations and the Full Cycle across Ethereum, XRPL, and Hedera, measured in seconds and shown in separate subplots for each platform.

### 5.1.1. Ethereum (did:ethr)

*Create* is implemented off–chain: a DID is implicitly derived from a newly generated address and instantiated locally (Figure 1a), so no transaction is submitted and consensus delay is avoided. This yields near–zero latency (mean $\approx 0.011$ s) as shown in Figure 4a and zero cost at creation time, at the expense of on–chain anchoring (immutability, timestamping) only beginning once a subsequent change is written. *Resolve* reconstructs the DID Document by reading ERC–1056 state and logs, potentially traversing historical events (Figure 2a), producing the highest resolve latency among the three platforms (mean $\approx 0.534$ s). *Update/Revoke/Delete* are smart–contract calls (e.g., setAttribute, revokeAttribute, changeOwner) that must be mined and confirmed (Figure 3a); means are ≈ 12.9 s (Update), 12.2 s (Revoke), and 12.6 s (Delete), closely tracking the 12.06 s block interval (Figure 5), with high variance and long tails (maxima up to ≈ 37 s).

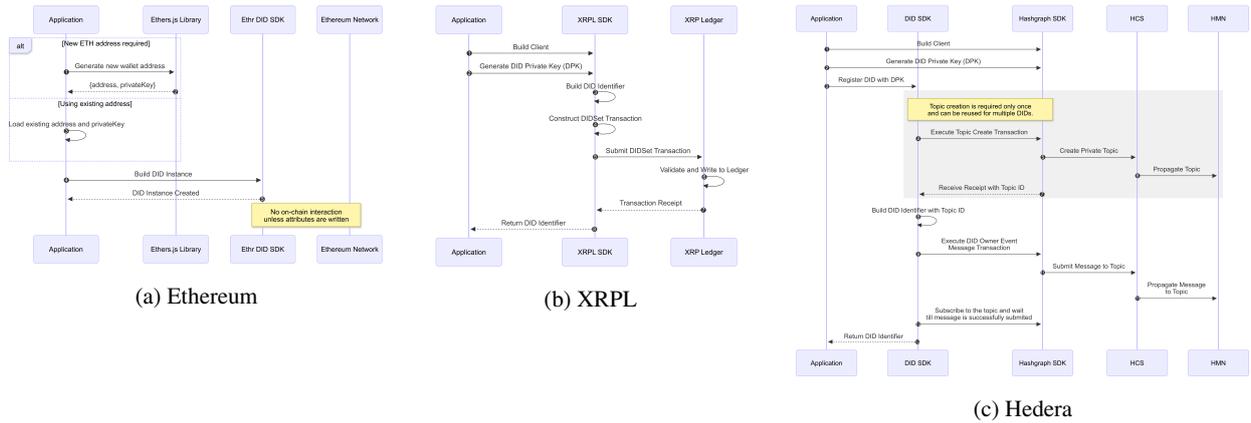

(a) Ethereum

(b) XRPL

(c) Hedera

**Figure 1:** DID Create operation sequence diagram.

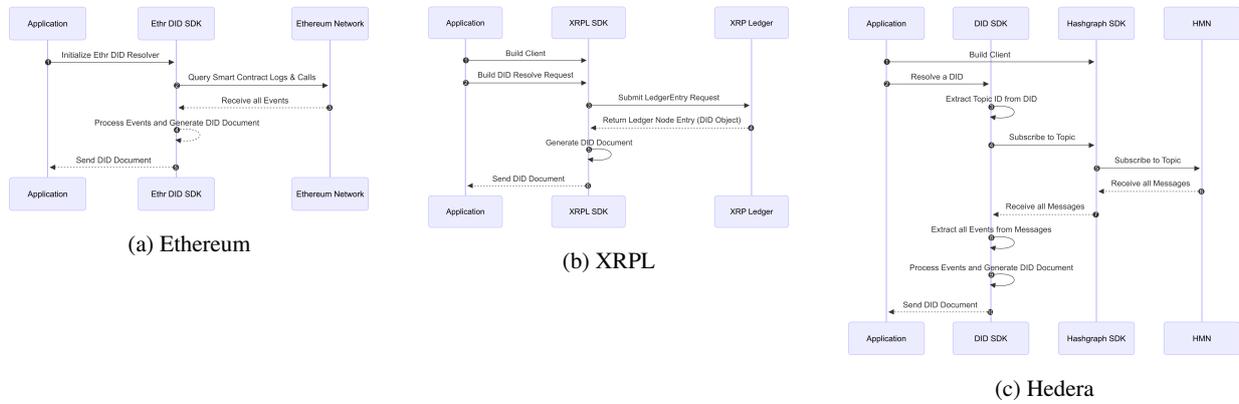

(a) Ethereum

(b) XRPL

(c) Hedera

**Figure 2:** DID Resolve operation sequence diagram.

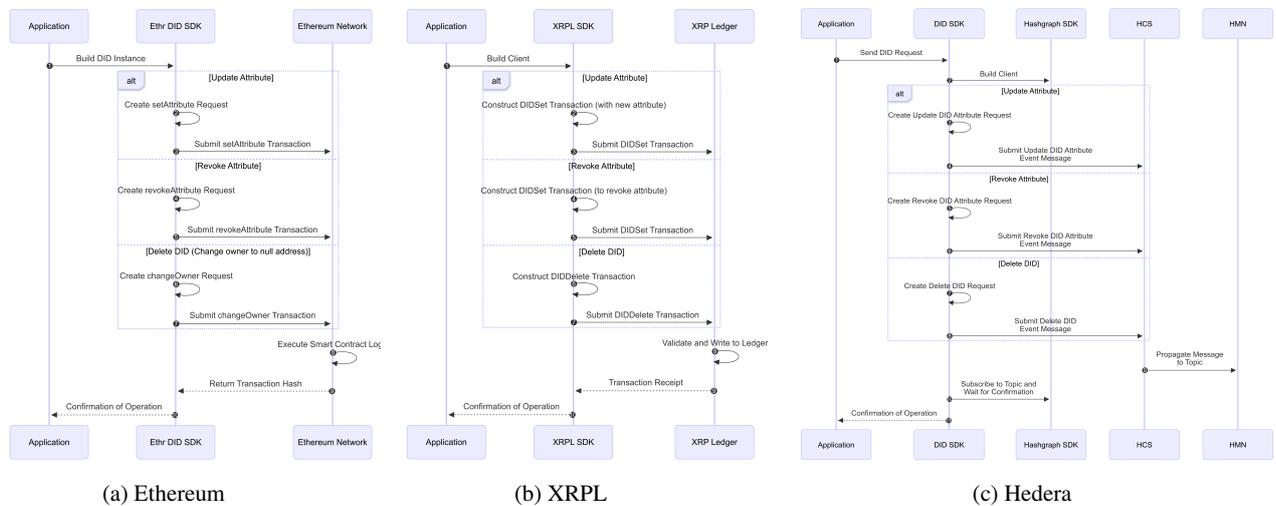

(a) Ethereum

(b) XRPL

(c) Hedera

**Figure 3:** DID Update, Revoke, and Delete operation sequence diagrams.

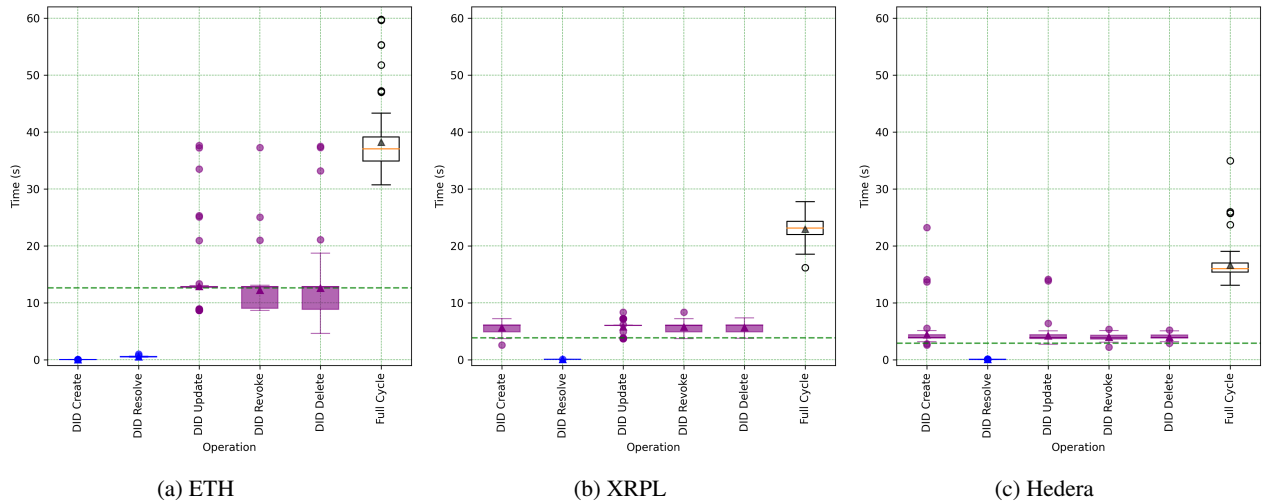

(a) ETH        (b) XRPL        (c) Hedera

**Figure 4:** Latency distributions for five DID operations and the Full Cycle across Ethereum, XRPL, and Hedera. For each operation, the boxplot summarizes the samples: the rectangle is the interquartile range (Q1–Q3), the horizontal line indicates the median, the triangular marker denotes the mean, and the whiskers extend to Q1-1.5×IQR and Q3+1.5×IQR; circles denote outliers beyond the whiskers (purple = on-chain operations, blue = off-chain, white = Full Cycle). The dashed green line marks each network's mean block/consensus interval.

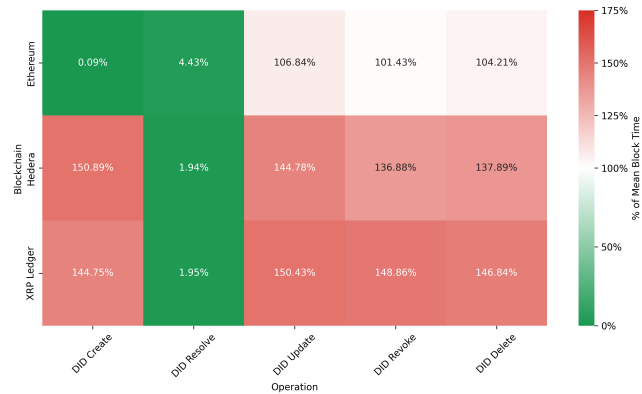

**Figure 5:** Relative latency heatmap for DID operations. Each cell reports (mean operation latency ÷ platform baseline interval) × 100%. Baseline block intervals: Ethereum 12.06 s, XRPL 3.87 s, and Hedera 2.90 s. Green denotes faster than one interval (<100%), white ≈ 100%, and red slower than one interval (>100%).

### 5.1.2. XRPL (did:xrpl)

*Create* requires a ledger write via `DIDSet` (Figure 1b), thus it waits for deterministic consensus; the mean latency is ≈ 5.6 s (Figure 4b), with low variance. *Resolve* performs a single `LedgerEntry` lookup of the stored DID object (Figure 2b), yielding a mean ≈ 0.076 s. *Update/Revoke/Delete* are on–ledger transactions (`DIDSet` for updates/revocations;

`DIDDelete` for deactivation) validated with deterministic finality (Figure 3b); means are ≈ 5.8 s (Update/Revoke) and ≈ 5.7 s (Delete), with tight dispersion. The relative latencies are about 145–150% of XRPL's block interval (Figure 5), reflecting modest overhead beyond base confirmation (transaction assembly, signature verification, and ledger serialization in the SDK/validator pipeline) while maintaining predictability due to XRPL's fixed-fee, deterministic consensus.

### 5.1.3. Hedera (did:hedera)

*Create* publishes an owner event to an HCS topic (Figure 1c); initializing a fresh topic adds overhead (≈ 2.5 s). In our setup, we create a topic once and reuse it for multiple DIDs; excluding topic creation, mean *Create* latency is ≈ 4.3 s, with higher variance and occasional outliers (up to ≈ 23 s) attributable to asynchronous propagation and mirror-node confirmation. *Resolve* subscribes to the topic via a mirror node and streams ordered DID messages to reconstruct state (Figure 2c), achieving the lowest mean (≈ 0.056 s) thanks to streaming APIs and the isolation of topic history. *Update/Revoke/Delete* appends signed messages to the DID's topic (Figure 3c), then confirms via the mirror-node pipeline; means are ≈ 4.2 s (Update), 4.0 s (Revoke), and 4.0 s (Delete), with occasional outliers (e.g., Update maximum ≈ 14.1 s). Overall, Hedera delivers the lowest absolute latencies for on-chain operations after topic setup (Figure 4c), and the relative latencies are about 137–150% of Hedera's consensus interval (Figure 5), reflecting modest overhead from message serialization, topic subscription, and mirror-node parsing beyond base consensus.

### 5.1.4. Non-obvious performance behaviors

Beyond confirming expected differences in absolute latency across platforms, a closer inspection of relative metrics and distributions reveals several non-obvious behaviors that are not apparent from raw averages alone.

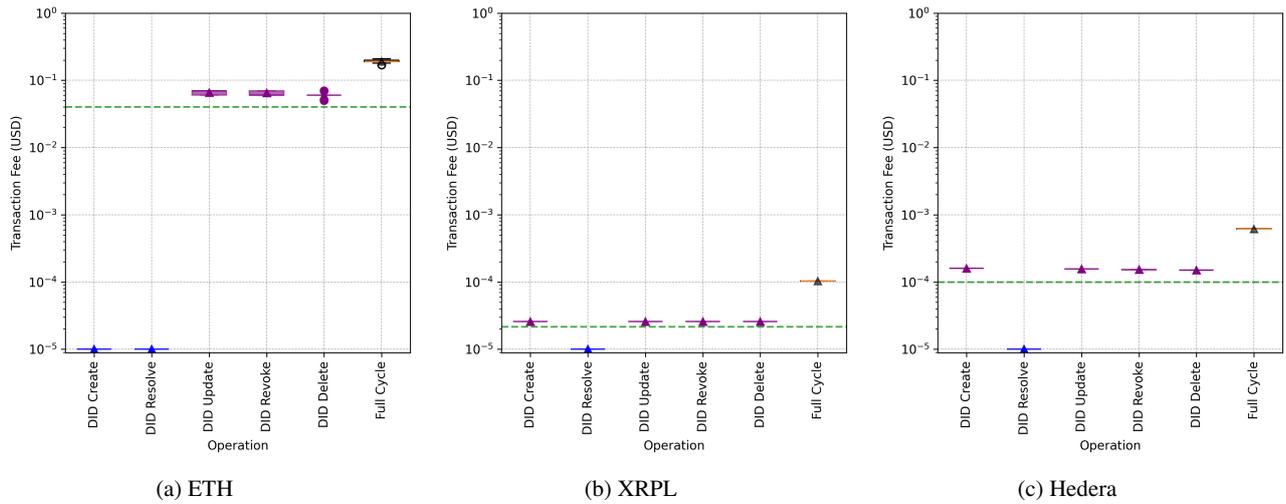

(a) ETH           (b) XRPL           (c) Hedera

**Figure 6:** Cost distributions of DID operations on Ethereum, XRPL, and Hedera (USD; priced on Apr. 15, 2025). The y-axis uses a logarithmic scale for cross-platform comparability. For each operation, the boxplot summarizes per-run network fees: the rectangle is the interquartile range (Q1–Q3), the horizontal line indicates the median, the triangular marker denotes the mean, and whiskers extend to Q1-1.5×IQR and Q3+1.5×IQR; circles denote outliers beyond the whiskers (purple = on-chain operations, blue = off-chain, white = Full Cycle). The dashed green line marks each network's mean native transfer fee used as a cost baseline. Off-chain operations have \$0.00 on-chain fee.

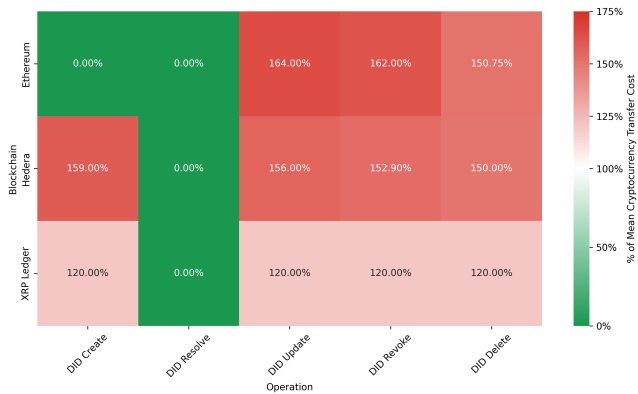

**Figure 7:** Relative transaction cost heatmap for DID operations. Each cell reports (mean operation fee ÷ platform's mean native transfer fee) × 100%. Mean native transfer fees: Ethereum \$0.04, XRPL \$0.000021, and Hedera \$0.0001. Green denotes cheaper than a simple value transfer (< 100%), white ≈ 100%, and red more expensive (> 100%). Off-chain operations have 0% on-chain fee.

First, relative latency highlights that for both XRPL and Hedera, DID operations consistently exceed a single native consensus interval (approximately 137–151% of baseline). This indicates that end-to-end latency is dominated not by consensus itself, but by client-side processing, SDK serialization, submission logic, and confirmation pipelines. Even on ledgers with deterministic and fast consensus, these non-consensus components emerge as first-order contributors to observed latency.

Second, Ethereum DID writes are largely block-time bound, with Update, Revoke, and Delete operations clustering around the block interval. However, Resolve operations show comparatively higher latency (mean ≈ 0.534 s), despite not requiring transaction inclusion. This overhead stems from DID state reconstruction via ERC-1056 storage and event logs, which often requires traversal of historical records. Thus, Ethereum's dominant non-obvious cost lies in resolution-time log and state access rather than consensus delay.

Third, while Hedera achieves low average latency, its tail behavior reveals occasional large outliers. These are attributable to asynchronous propagation and mirror-node confirmation rather than consensus delays. This highlights a practical deployment insight: in event-stream architectures, read and confirmation infrastructure can become the bottleneck even when consensus is fast.

Finally, the results expose a clear predictability–speed trade-off across platforms. XRPL is not the fastest on average, but it exhibits the tightest latency distributions and the most deterministic behavior. Hedera offers lower mean latency with occasional spikes, while Ethereum shows longer tails. This distinction suggests that predictability and stability may be as important as raw speed when selecting a ledger for DID deployments.

### 5.2. Transaction Cost Analysis

Figure 6 presents the cost distributions of individual DID operations and the Full Cycle across Ethereum, XRPL, and Hedera. Costs are reported in USD using market prices on Apr. 15, 2025. Each subplot overlays a dashed green line representing the mean cost of a simple native transfer on that network (Ethereum: \$0.04 at 1.2 Gwei; XRPL: \$0.000021;

Hedera: $0.0001), providing a common baseline for comparison. For Ethereum, we fixed the gas price at 1.2 Gwei for both baseline transfers and ERC-1056 calls to remove gas-price confounding. Detailed statistics appear in Appendix Table 8.

### 5.2.1. Ethereum

As shown in Figure 6a, on-chain DID operations (Update, Revoke, Delete) average $0.066, $0.065, and $0.060, respectively, while *Create* and *Resolve* incur $0.00 on-chain fees because did:ethr creation is off-chain and resolution is read-only. Relative to the $0.04 transfer baseline, these write operations are ∼ 1.50–1.64× the baseline (Figure 7). The premium stems from contract execution: state writes and event emissions in ERC-1056 consume storage and log indexing, and signature verification adds compute beyond a simple ETH transfer.

### 5.2.2. XRPL

Figure 6b shows uniformly low, fixed fees for `DIDSet`/`DIDDelete`; our runs priced at $0.000026 per transaction, with *Resolve* at $0.00. The baseline used in the plots (dashed green line) is a standard payment fee of $0.000021 (10 drops) [35], so DID writes are ≈ 120% of baseline (see Figure 7). The fees are ledger-scheduled and deterministic, producing essentially zero dispersion and highly predictable costs.

### 5.2.3. Hedera

On Hedera (Figure 6c), *Create* averages $0.00016 and *Update/Revoke/Delete* $0.00015; the Full Cycle totals $0.00062. Compared to the $0.0001 `CryptoTransfer` baseline [36], these operations are ∼ 1.5–1.6× the baseline (Figure 7). Although DID writes use `ConsensusSubmitMessage`, the fee schedule accounts for payload bytes and signature handling; the structured JSON payloads required by the method and SDK introduce a moderate premium relative to a basic transfer.

### 5.2.4. Non-obvious cost dynamics

Beyond differences in absolute fees, the relative cost analysis reveals a counter-intuitive pattern across platforms. Despite operating under vastly different fee models and absolute cost levels, both Ethereum and Hedera exhibit a remarkably similar relative overhead for DID write operations. Across Update, Revoke, and Delete, DID transactions incur approximately 1.5–1.6× the cost of a native value transfer on both platforms. This suggests that the economic premium introduced by DID-specific logic is comparable across fundamentally different ledger architectures, and that relative overhead is driven more by operation complexity than by the underlying fee market.

In contrast, XRPL displays a distinct cost profile. DID operations on XRPL incur a much lower relative overhead, remaining close to 120% of a baseline transfer, with zero variance across runs. This reflects the ledger-native DID object model, where identity updates are handled as first-class state changes rather than layered on top of general-purpose

**Table 4**
MLS for DID Operations

(a) DID Update transactions

| Chain | Bits/Token | Avg Tokens | Bits/Txn |
|---|---|---|---|
| Ethereum | 0.0034 | 24.0 | 0.082 |
| Hedera | 0.0037 | 19.0 | 0.071 |
| XRPL | 0.0024 | 36.0 | 0.088 |

(b) DID Revoke transactions

| Chain | Bits/Token | Avg Tokens | Bits/Txn |
|---|---|---|---|
| Ethereum | 0.0036 | 23.0 | 0.083 |
| Hedera | 0.0040 | 17.0 | 0.068 |
| XRPL | 0.0024 | 35.0 | 0.083 |

(c) DID Delete transactions

| Chain | Bits/Token | Avg Tokens | Bits/Txn |
|---|---|---|---|
| Ethereum | 0.0035 | 22.0 | 0.078 |
| Hedera | 0.0038 | 17.0 | 0.064 |
| XRPL | 0.0020 | 40.0 | 0.079 |

execution or messaging frameworks. Together, these results show that absolute cost alone is an incomplete indicator of DID efficiency, and that relative overhead exposes architectural differences that are otherwise obscured.

## 5.3. Privacy Analysis

### 5.3.1. Operation–specific leakage

Table 4 reports three metrics per chain and operation: bits per token, average tokens per transaction, and total bits per transaction. Hedera shows the highest bits–per–token (more variable fields), Ethereum is slightly lower, and XRPL is lowest (more uniform fields). Token volume reverses this ordering: XRPL payloads are the most verbose (≈ 35–40 tokens), Ethereum is mid–range (≈ 22–24), and Hedera is leanest (≈ 17–19). Combined, XRPL yields the largest total leakage for *Update* (0.088 bits/txn) and *Delete* (0.079), and roughly matches Ethereum on *Revoke* (0.083). Hedera is consistently lowest: 0.070 (Update), 0.068 (Revoke), 0.064 (Delete).

### 5.3.2. Aggregate leakage across all operations

Table 5 aggregates across Update, Revoke, and Delete. Individual operations fall in the 0.06–0.09 bits/txn range; aggregated MLS is 0.20–0.25 bits/txn, indicating non–trivial, observable variability in all on–chain payloads. Hedera has the lowest aggregate MLS (0.20; 60 total bits across 300 txns), Ethereum is intermediate (0.24; 72 bits), and XRPL highest (0.25; 75 bits), primarily due to XRPL's larger token counts despite its lower per–token entropy. Within each chain, *Delete* leaks least, followed by *Update*, then *Revoke*; differences are modest. Because MLS correlates with fingerprintability, Hedera's lower scores suggest a slightly stronger privacy posture; XRPL's verbosity increases linkage risk; Ethereum sits between.

**Table 5**
Total MLS across all operations

| Chain | Bits/Token | Tokens/Txn | Bits/Txn | Bits (Total) |
|-------|-----------|-----------|----------|--------------|
| Ethereum | 0.0035 | 23 | 0.24 | 72 |
| Hedera | **0.0038** | **17.7** | **0.20** | 60 |
| XRPL | **0.0023** | 37 | 0.25 | 75 |

### 5.3.3. Interpreting MLS differences

Observed MLS differences arise from two orthogonal design factors: *payload verbosity* (average tokens per transaction) and *field variability* (bits per token). For example, XRPL exhibits the lowest bits-per-token, reflecting highly standardized and rigid transaction schemas; however, its comparatively verbose payloads (35–40 tokens) inflate total bits per transaction, yielding higher MLS overall. Hedera shows the opposite pattern: slightly higher bits-per-token due to more flexible fields, but substantially leaner payloads (17–19 tokens), resulting in the lowest total leakage across all operations. Ethereum occupies an intermediate position on both dimensions. Across operations, *Delete* consistently leaks least because it minimizes optional fields and structural variation, whereas *Revoke* and *Update* expose more configuration-dependent metadata.

### 5.3.4. Practical implications of MLS magnitudes

While MLS values per operation (e.g., 0.06–0.09 bits/txn) are small, they accumulate across repeated interactions and can aid linkage when combined with auxiliary knowledge (known accounts, timing, endpoint patterns). Under independence assumptions, $b$ bits narrow an adversary's effective anonymity set to roughly $2^b$ candidates (approximate, not a guarantee). For example, an actor performing $N = 100$ updates at 0.08 bits each yields $\approx$ 8 bits in aggregate ($\sim$ 256 equivalence classes). Even weaker per–op signals become actionable when correlated over time (cadence of key rotations, recurring service endpoint structures, topic/account reuse) or joined across datasets. Thus, MLS highlights which fields drive fingerprintability and motivates concrete mitigations (schema normalization, endpoint hashing/referencing, minimizing optional fields, batching predictable changes), even when absolute bit values per transaction appear modest.

### 5.3.5. Conceptual linkage scenario

Although we do not perform a full linkage-attack evaluation, we sketch how an adversary could exploit metadata variability in practice. A passive observer can cluster operations by their tokenized metadata fields (e.g., service-endpoint formats, key rotation patterns, transaction-structure signatures). Higher MLS indicates more distinguishable payloads and therefore a larger feature space for clustering, whereas lower MLS constrains the observer to near-uniform records.

For example, if two operations share similar token distributions under a high-entropy schema, an attacker can assign higher likelihood that they originate from the same DID controller. Conversely, chains with very low MLS (e.g., Hedera in our results) expose fewer discriminative fields and generate tighter clusters, reducing linkage confidence. This conceptual analysis shows how MLS relates to practical fingerprintability.

## 5.4. Interoperability in DID Systems

Interoperability is the ability of heterogeneous DID methods and platforms to resolve, interpret, and verify each other's identifiers and credentials. It depends on adherence to shared specifications and data models so that a DID issued on one network can be reliably resolved and used on another. A key enabler is the `Universal Resolver (UR)` [54], an open–source DIF [55] service that loads method–specific drivers behind a common API; methods with production UR drivers are more readily consumable by standards compliant wallets, verifiers, and agents. Interoperability also hinges on community governance and ongoing participation in standards bodies.

### 5.4.1. Comparative analysis of interoperability

*Ethereum-DID.* Ethereum-based DID methods are among the most mature and widely deployed W3C–aligned approaches, benefiting from Ethereum's global developer base and a large ecosystem of SSI tooling. In practice, Ethereum DIDs are commonly integrated through modular agent frameworks such as Veramo [49], and are resolvable via the Universal Resolver, enabling resolver–based interoperability across wallets and verifiers. Beyond core standards alignment and active participation in DID standardization communities, Ethereum's strongest interoperability advantage is ecosystem breadth: Alchemy's Dapp Store catalog alone lists 46 decentralized identity tools on Ethereum, spanning wallets, credential infrastructure, reputation systems, compliance/KYC services, and developer SDKs, indicating dozens of actively maintained commercial and open-source projects in this space [56].

*Hedera-DID.* Registered in the W3C DID Method Registry [57], Hedera's DID methods anchor DID documents via the HCS. Hedera also maintains visible engagement in identity standardization and open–source governance through the DIF and the Linux Foundation Decentralized Trust (LFDT), where the Hedera codebase and related ecosystem work are developed under the Hiero umbrella [58]. On the credential layer, the Hiero AnonCreds Method [59] enables privacy–preserving AnonCreds objects to be written to and resolved from the ledger, which supports practical interoperability with widely used SSI stacks such as Hyperledger Aries. Developer-facing tooling further strengthens this pathway: the Hiero DID SDK JavaScript [52] provides concrete APIs for creating and resolving Hedera DIDs and working with AnonCreds resources in application workflows. Finally, with a fully implemented UR driver, Hedera DIDs can be resolved using standard resolver infrastructure, enabling standards–compliant, resolver–based interoperability with wallets, agents, and verifiers across the broader DID ecosystem.

**Table 6**
Interoperability Comparison of DID Methods

| Factor | Ethereum DID | Hedera DID | XRPL DID |
|---|---|---|---|
| **W3C DID Core Compliance** | Yes | Yes | Yes |
| **Standards Participation** | DIF, W3C | DIF, W3C, LFDT | Aligns with W3C; not a DIF member |
| **Universal Resolver Support** | Full | Full | Planned (driver in progress) |
| **VC Format Support** | JSON-LD, JWT | JSON-LD, AnonCreds via Aries | JSON-LD; XRPL Credentials (in progress) |
| **Governance Model** | Open, Ethereum consensus | Hedera Council; open-sourced under LF | Validator governance; RippleX-led proposals |
| **Cross-Chain Potential** | Strong (multi-chain EVM) | Emerging (SSI/Aries integration) | Early, with pilots under exploration |
| **Commercial Identity Projects** | SpruceID [37]<br>BrightID [38]<br>Proof of Humanity [39]<br>Fractal ID [40] | Meeco [41]<br>EarthID [42]<br>IDTrust [43]<br>DSR Corporation [44] | idOS [45]<br>Sologenic [46]<br>Redimi [47] |
| **Open-source Identity Projects** | walt.id [48]<br>Veramo [49] | ACA-Py [50]<br>Credo [51]<br>Hiero [52] | XRPL Credentials [53] |

*XRPL-DID.* XRPL introduced a native, W3C–conformant DID method through the XLS–40d standard. While broader resolver–based interoperability (e.g., Universal Resolver integration) depends on the continued maturation of drivers and ecosystem tooling, the method is explicitly structured to fit global DID resolution conventions as adoption increases. In parallel, the XRPL community emphasizes identity for compliance–oriented and regulated use cases: the evolving *Credentials* (XLS-70 [53]) feature provides on-ledger primitives for authorization and compliance workflows while aiming to preserve privacy and decentralization principles [60].

*Summary.* All three methods conform to the W3C DID Core specification but differ in tooling maturity and ecosystem adoption (Table 6). Ethereum benefits from early adoption and established UR support. Hedera combines strong standards participation with mature SSI integrations and fully implemented UR support. XRPL is newer but was designed with W3C interoperability in mind, with active development toward resolver integration and cross–ledger support. Continued deployment of UR drivers and broader cross–ecosystem credential verification will be pivotal for achieving seamless, global interoperability.

# 6. Discussion

## 6.1. Implications for Practitioners and Researchers

The results provide evidence-based guidance for system selection and configuration (RQ3). No single ledger optimizes latency, cost, and privacy simultaneously; instead, each platform embodies distinct trade-offs that should be matched to application requirements.

Hedera achieves the lowest on-chain latency and the smallest metadata leakage, making it well suited for latency- and privacy-sensitive workflows such as interactive credential exchanges or frequent key rotation. XRPL offers highly predictable performance and consistently low, fixed fees, which favors large-scale and cost-sensitive deployments, although its more verbose transaction payloads increase metadata exposure. Ethereum trades higher cost and slower on-chain writes for mature tooling, programmability, and deep integration with the broader Web3 ecosystem, which can be decisive when extensibility, composability, and interoperability are primary concerns.

For researchers, this study provides a unified and reproducible benchmarking framework, a cross-platform evaluation methodology, and an analytical model for quantifying on-chain metadata exposure. These artifacts enable systematic comparison of DID implementations and support further empirical work on performance, cost, and privacy trade-offs in decentralized identity systems.

## 6.2. Threats to Validity
### 6.2.1. Scope

This study evaluates three production-grade, W3C-aligned DID methods that represent distinct ledger architectures and are implemented via public SDKs on testnets. While this limits direct claims about scalability under high load, security guarantees, or operational behavior in production deployments, it enables controlled and repeatable measurement of system-level properties intrinsic to ledger architecture and DID method design. The observed differences in latency structure, cost overhead, resolution behavior, and metadata exposure stem from architectural choices that are largely independent of network scale or adversarial conditions. Accordingly, the results characterize comparative design trade-offs rather than absolute performance at scale, and may not generalize to other DID ecosystems (e.g., ION, PolygonID, Solana-based methods) or deployment contexts such as enterprise or permissioned environments.

### 6.2.2. Testnet vs. Mainnet

All experiments were run on public testnets to ensure reproducibility. Testnets do not reflect mainnet conditions such as congestion, gas–price volatility, or transaction queuing, which are especially relevant for Ethereum. As a result, absolute latency and cost values likely underestimate real-world performance. However, because our analysis also explores *relative* differences driven by architectural design and SDK workflows, we expect the comparative ordering of platform behaviors to remain stable. Mainnet validation is left for future work.

### 6.2.3. SDK variability

Results depend on specific SDK versions and default configurations. Different client libraries or future updates may alter latency, serialization behavior, or payload structure. Our findings therefore reflect the reference implementations evaluated rather than immutable platform characteristics.

### 6.2.4. Sampling

We collected 100 samples per operation, which captures central tendencies but does not model rare tail events or apply formal statistical inference. Larger samples or stress scenarios could strengthen confidence intervals around our estimates.

## 7. Conclusion

This paper presented a comparative, empirical study of DID systems on three prominent distributed ledger platforms—Ethereum, Hedera, and XRPL. Using reference SDKs and a unified benchmarking framework, we measured latency, transaction cost, and on-chain metadata exposure, and linked these outcomes to concrete design choices: off-vs. on-chain instantiation, contract-based vs. ledger-native vs. event-stream transaction models, and resolver workflows.

*Key findings.* Our evaluation reveals non-obvious system-level insights that extend beyond absolute performance differences. First, end-to-end DID latency is strongly influenced by non-consensus components such as SDK processing, submission logic, and confirmation infrastructure. Resolution paths are highly method-specific and exhibit distinct latency characteristics, with log and state traversal on Ethereum, direct keyed ledger lookups on XRPL, and topic replay via mirror nodes on Hedera. This makes read-path design as critical as write performance. Second, relative cost analysis exposes a counter-intuitive result: despite vastly different fee models, Ethereum and Hedera incur a similar relative premium for DID write operations, indicating that operation complexity, rather than the underlying fee market, drives this overhead. XRPL departs from this pattern by maintaining lower and more stable relative costs due to its ledger-native DID model. Third, privacy leakage is driven by different mechanisms across platforms. XRPL exhibits higher aggregate leakage primarily due to payload verbosity, whereas Ethereum and Hedera expose fewer tokens despite higher per-token variability, demonstrating that metadata exposure depends on payload structure as well as entropy.

*Design implications.* Taken together, these findings indicate that DID performance and privacy are shaped not only by headline ledger properties and also by end-to-end system design. Where work is placed (off-chain vs. on-chain), how state is reconstructed (logs, keyed reads, or replay), and how client and indexing infrastructure is integrated all emerge as first-order factors. A practical design pattern that follows is to anchor constant-size commitments on-chain, keep encrypted, content-addressed DID Documents off-chain, and model each DID as an append-only event log supporting ordered updates, key rotation, and recovery. This approach preserves low and predictable latency and fees while constraining the observable metadata surface.

*Limitations and future work.* Our study focuses on system-level metrics measured under controlled conditions and does not capture all deployment contexts. Future work should extend this analysis along complementary dimensions, including scalability under concurrent DID workloads, security and adversarial resilience, and validation on mainnet deployments under realistic congestion and fee dynamics. In addition, the evaluation should be broadened to include a wider range of DID methods and technology stacks.

In sum, our measurements and analysis demonstrate how architectural and infrastructural choices, often outside the core consensus protocol, translate into concrete latency, cost, and metadata exposure properties of DID operations, providing actionable guidance for selecting and configuring DID stacks in real-world deployments.

## A. My Appendix

### A.1. Replication Package and Data Availability

To support transparency and reproducibility, we provide a complete replication package accompanying this paper. It includes the benchmarking scripts, raw result files, analysis notebooks, and documentation required to reproduce all figures and tables. The package is publicly available at: `https://anonymous.4open.science/r/DID-Research-C147`. All materials are released under an open-source license and are compatible with the experimental methodology described in Section 4.

### A.2. Summary Statistics

**Table 7**
Summary statistics of DID operation latencies (in seconds) by DLT.

|  | Metric | DID Create | DID Resolve | DID Update | DID Revoke | DID Delete | Full Cycle |
|---|---|---|---|---|---|---|---|
| Ethereum | Mean | 0.011 | 0.534 | 12.885 | 12.232 | 12.567 | 38.230 |
|  | Std | 0.005 | 0.063 | 4.931 | 3.488 | 4.815 | 7.387 |
|  | Min | 0.010 | 0.423 | 8.670 | 8.685 | 4.628 | 30.701 |
|  | Max | 0.015 | 0.966 | 37.643 | 37.280 | 37.474 | 63.729 |
| XRPL | Mean | 5.602 | 0.076 | 5.821 | 5.761 | 5.683 | 22.943 |
|  | Std | 1.183 | 0.006 | 1.197 | 1.199 | 1.081 | 2.221 |
|  | Min | 2.594 | 0.075 | 3.729 | 3.708 | 3.743 | 16.183 |
|  | Max | 7.229 | 0.078 | 8.347 | 8.356 | 7.342 | 27.739 |
| Hedera | Mean | 4.375 | 0.056 | 4.199 | 3.970 | 3.999 | 16.599 |
|  | Std | 2.396 | 0.008 | 1.504 | 0.537 | 0.441 | 2.820 |
|  | Min | 2.570 | 0.051 | 2.698 | 2.225 | 2.875 | 13.091 |
|  | Max | 23.192 | 0.100 | 14.128 | 5.328 | 5.231 | 34.932 |

**Table 8**
Summary statistics of DID operation costs (in USD) by DLT.

|  | Metric | DID Create | DID Resolve | DID Update | DID Revoke | DID Delete | Full Cycle |
|---|---|---|---|---|---|---|---|
| Ethereum | Mean | 0.000000 | 0.000000 | 0.065600 | 0.064800 | 0.060300 | 0.190700 |
|  | Std | 0.000000 | 0.000000 | 0.004989 | 0.005021 | 0.005214 | 0.008791 |
|  | Min | 0.000000 | 0.000000 | 0.060000 | 0.060000 | 0.050000 | 0.170000 |
|  | Max | 0.000000 | 0.000000 | 0.070000 | 0.070000 | 0.070000 | 0.210000 |
| XRPL | Mean | 0.000026 | 0.000000 | 0.000026 | 0.000026 | 0.000026 | 0.000104 |
|  | Std | 0.000000 | 0.000000 | 0.000000 | 0.000000 | 0.000000 | 0.000000 |
|  | Min | 0.000026 | 0.000000 | 0.000026 | 0.000026 | 0.000026 | 0.000104 |
|  | Max | 0.000026 | 0.000000 | 0.000026 | 0.000026 | 0.000026 | 0.000104 |
| Hedera | Mean | 0.000159 | 0.000000 | 0.000156 | 0.000153 | 0.000150 | 0.000618 |
|  | Std | 0.000000 | 0.000000 | 0.000001 | 0.000001 | 0.000000 | 0.000001 |
|  | Min | 0.000159 | 0.000000 | 0.000156 | 0.000152 | 0.000150 | 0.000617 |
|  | Max | 0.000159 | 0.000000 | 0.000156 | 0.000154 | 0.000150 | 0.000619 |

### A.3. Hedera DID Update Transaction Metadata

```
{
  "transaction_info": {
    "initial_transaction_id": {
      "account_id": "0.0.5436919",
      "nonce": 0,
      "scheduled": false,
      "transaction_valid_start": "1747825185.015271329"
    },
    "number": 1,
    "total": 1
  },
  "consensus_timestamp": "1747825198.342924000",
  "message": {
    "message": {
      "timestamp": "2025-05-21T10:59:57.9052",
      "operation": "update",
      "did": "did:hedera:testnet:zEqgDR8Fh2ZXbuLQTi1Bc_0.0.5649399",
      "event": {
        "Service": {
          "id": "did:hedera:testnet:zEqgDR8Fh2ZXbuLQTi1Bc_0.0.5649399#service-0",
          "type": "LinkedDomains",
          "serviceEndpoint": "https://example.com/0"
        }
      }
    },
    "signature": "hdcVRQRQgWHs1B8SOHhTebId0Yi1cB/YTh8fuwmbxcvMaJTzMzjMucKJ15=="
  },
  "payer_account_id": "0.0.5436919",
  "running_hash": "kDGNsUl2nWzC29eOMGWT!tUv1LBE2xpucptK6hzCXOxOMGxoahpxIi6EhvY",
  "running_hash_version": 3,
  "sequence_number": 818,
  "topic_id": "0.0.5649399"
}
```

This payload is the mirror-node record of a single Hedera HCS DID Update. In the MLS analysis, we treat these observable fields as tokens for entropy calculation.

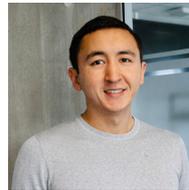

Abylay Satybaldy received the Ph.D. degree in Computer Science from the Norwegian University of Science and Technology (NTNU). He is currently a Senior Research Associate at Exponential Science. His research focuses on decentralized identity (DID/SSI), blockchain performance and benchmarking, and empirical analysis of decentralized finance systems. He has contributed to open-access datasets, analytical frameworks, and industry research reports, and actively participates in research communities related to decentralized identity and distributed ledger technologies.

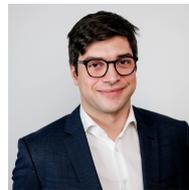

Kamil Tylinski is a Ph.D. candidate in Computer Science at University College London (UCL), where his research focuses on quantitative analysis of trust, explainability, transparency, and robustness in machine learning and distributed ledger technologies. He is currently the Head of Data and Artificial Intelligence at Exponential Science Foundation. His work spans applied AI/ML, NLP, and data analytics across academia, industry, and policy-oriented research. He has led and contributed to research projects, analytical frameworks, and industry reports in AI and DLT, and actively engages in academic supervision, interdisciplinary research related to artificial intelligence, explainability of AI models, DLT, and emerging technologies.

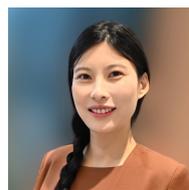

Jiahua Xu is an Associate Professor in Financial Computing and Programme Director of the MSc Emerging Digital Technologies at UCL. Her research focuses on blockchain economics and decentralized finance, with publications in Usenix Security, ACM IMC, ACM ASIACCS, FC, IEEE ICDCS and IEEE COMST.